\documentclass[fleqn,usenatbib]{mnras}

\usepackage{newtxtext,newtxmath}

\usepackage[T1]{fontenc}
\usepackage{ae,aecompl}

\usepackage{graphicx}	
\usepackage{amsmath}	
\usepackage{amssymb}	

\title[The SEDs of high-redshift LBGs]{Probing Cosmic Dawn: Modelling the Assembly History, SEDs, and Dust Content of Selected $z\sim9$ Galaxies}

\author[H. Katz et al.]{Harley Katz$^{1}$\thanks{E-mail: harley.katz@physics.ox.ac.uk},
Nicolas Laporte$^{2}$, Richard S. Ellis$^{2}$, Julien Devriendt$^{1}$, \newauthor \& Adrianne Slyz$^{1}$
\\
$^{1}$Astrophysics, University of Oxford, Denys Wilkinson Building, Keble Road, Oxford OX1 3RH, UK \\
$^{2}$Department of Physics \& Astronomy, University College London, London, WC1E 6BT, UK
}

\date{Accepted XXX. Received YYY; in original form ZZZ}

\pubyear{2015}

\begin{document}
\label{firstpage}
\pagerange{\pageref{firstpage}--\pageref{lastpage}}
\maketitle

\begin{abstract}
The presence of spectroscopically confirmed Balmer breaks in galaxy spectral energy distributions (SEDs) at $z>9$ provides one of the best probes of the assembly history of the first generations of stars in our Universe.  Recent observations of the gravitationally lensed source, MACS~1149\_JD1 (JD1), indicate that significant amounts of star formation likely  occurred at redshifts as high as $z\simeq15$.  The inferred stellar mass, dust mass, and assembly history of JD1, or any other galaxy at these redshifts that exhibits a strong Balmer break, can provide a strong test of our best theoretical models from high-resolution cosmological simulations.  In this work, we present the results from a cosmological radiation-hydrodynamics simulation of the region surrounding a massive Lyman-break galaxy. For two of our most massive systems, we show that dust preferentially resides in the vicinity of the young stars thereby increasing the strength of the measured Balmer break such that the simulated SEDs are consistent with the photometry of JD1 and two other $z>9$ systems (GN-z10-3 and GN-z9-1) that have proposed Balmer breaks at high redshift.  We find strong variations in the shape and luminosity of the SEDs of galaxies with nearly identical stellar and halo masses, indicating the importance of morphology, assembly history, and dust distribution in making inferences on the properties of individual galaxies at high redshifts.  Our results stress the importance that dust may play in modulating the observable properties of galaxies, even at the extreme redshifts of $z>9$. 
\end{abstract}




\section{Introduction}
The quest for cosmic dawn, the epoch corresponding to the emergence of the first galaxies from the dark, neutral intergalactic medium produced at recombination, represents the final observational frontier in constructing a physical picture of galaxy evolution. As the first generation of stars were devoid of heavy elements, a potential observational strategy for identifying such objects would be to use the James Webb Space Telescope (JWST) to locate sources with spectra lacking metallic nebular emission lines or that exhibit intense ionised helium, characteristic of hot stars free from line blanketing \citep{Schaerer2002}.  However, numerical simulations suggest that such low mass halos are likely rapidly self-enriched to metallic abundances of $\simeq10^{-3}Z_{\odot}$ by early supernovae \citep{Smith2015}, indicating that such ``Population III" galaxies would be short-lived and consequently difficult to find.

A more practical strategy is to link cosmic dawn to the beginnings of cosmic reionization - the transformation of neutral hydrogen in the intergalactic medium (IGM) to an ionised state. Although uncertainties remain, it seems likely that reionization was initiated by star-forming galaxies (see \citealt{Stark2016} for a review). The most recent analysis of Thomson scattering in the IGM from measures of polarisation and temperature fluctuations in the microwave background \citep{Planck2018} suggests reionization began relatively late, corresponding to redshifts of $z\sim 10-15$. An intriguing signal of 21cm absorption in the microwave background has been claimed corresponding to Lyman~$\alpha$ emission from early sources at $z\simeq15$ \citep{Bowman2018}.

Unfortunately, current observational facilities are not capable of detecting early galaxies beyond a redshift of $z\simeq11-12$ where both continuum and nebular emission are expected to be predominantly in the near-infrared. The Hubble Space Telescope (HST) has a wavelength range limited to $\lambda<1.6\mu$m, while the near-infrared performance of large ground-based telescopes is limited by the thermal sky background. The most distant spectroscopically-confirmed sources are at $z=9.11$ \citep{Hashimoto2018} and $z\simeq11$ \citep{Oesch2016}.  

A potential way to provide a first estimate of when cosmic dawn occurred would be to estimate the age and dust content of galaxies at slightly lower redshift, e.g. $z\simeq8-10$. As cosmic time is compressed with respect to redshift at early times, even an approximate stellar age at $z\simeq9$, when the Universe was 550~Myr old, can provide a meaningful indication of when the first generation of stars emerged.  \cite{Hashimoto2018} provided an illustration of this method by analysing the SED, dust content, and line emission from a gravitationally-lensed source MACS~1149\_JD1 (hereafter JD1) at $z=9.11$. Their analysis indicated that a significant fraction of the stars in JD1 likely formed at a redshift $z_{\rm form}\simeq15.4\pm2.3$. Likewise \cite{Laporte2017} estimated a dust mass of $6\times10^6{\rm M_{\odot}}$ for a gravitationally-lensed source at a redshift $z=8.38$ and claimed that such a mass, coupled with estimates of the past star formation rate, could be used to pinpoint the likely period when chemical enrichment began.

Inferring the past assembly history of $z\simeq8-10$ galaxies to estimate when cosmic dawn occurred might be the only promising route while we await JWST, but the method has many uncertainties. Of particular interest is whether the stellar and dust masses, and the inferred assembly histories of these observed galaxies, are reasonable in the context of models of hierarchical structure formation in the standard $\Lambda$ cold dark matter cosmology. \cite{Hashimoto2018} claimed that the bulk of the stars observed in JD1 at $z=9.11$ formed at $z\simeq12-15$ after which the star formation rate must have declined with time to match a prominent Balmer break seen in the SED.  Star formation rates that decline with time at $z\simeq10-15$ contrast with high-resolution cosmological simulations which mostly predict bursty star formation histories that are generally rising with time \citep[e.g.][]{Rosdahl2018}.

\begin{figure*}
\includegraphics[width=15cm]{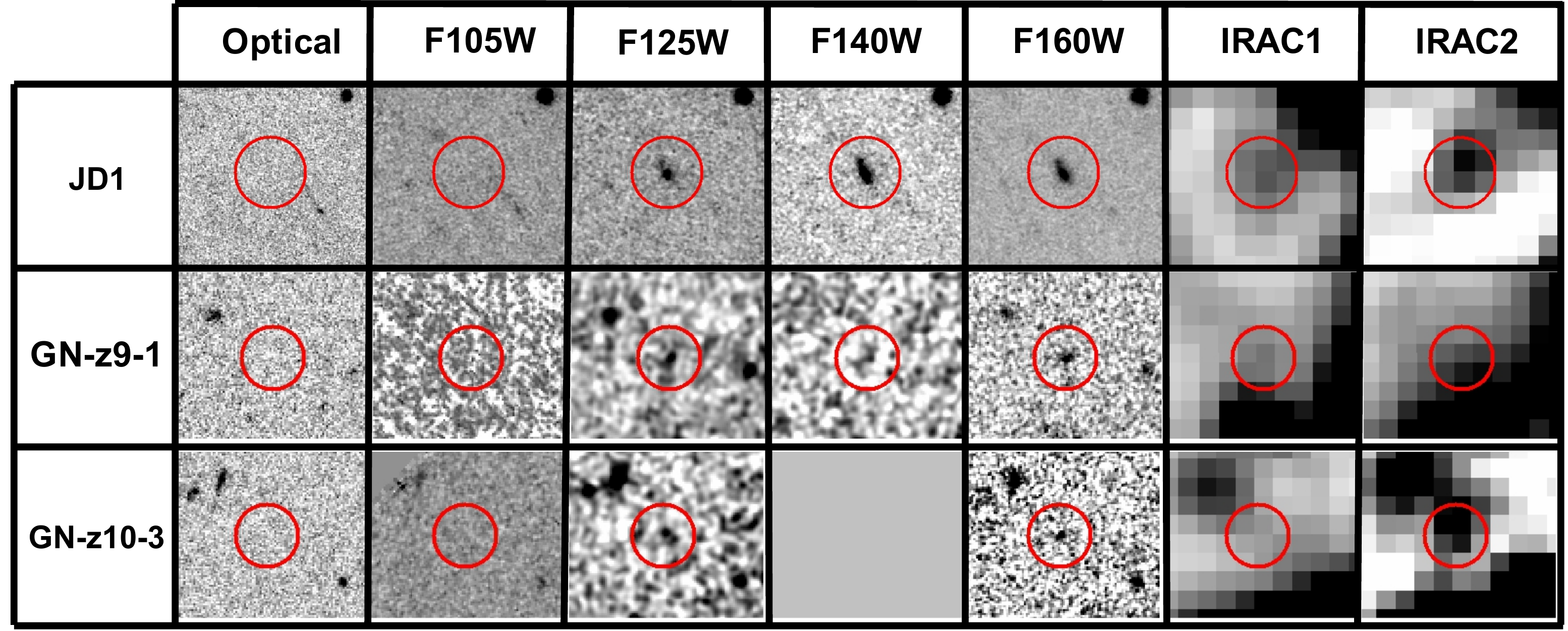}
\caption{ Stamps of the three high-redshift candidates discussed in this paper. The size of each stamp is 5''$\times$5'' and the position of each target is displayed by a red 1'' radius circle. For JD1, the optical stamps is the sum of F435W, F606W and F814W whereas for GN-z10-3 and GN-z9-1 it corresponds to a sum of F814W and F850LP. }
\end{figure*}

Comparing simulated galaxies with observations is highly non-trivial as massive systems are likely to be dust and metal-enriched and the distribution of stars, gas and dust can significantly alter the intrinsic spectrum via absorption and scattering.  Analytic models demonstrate that the distribution of dust in a galaxy is key to reproducing the spectral properties of observed galaxies \citep[e.g.][]{Charlot2000}.  The importance of dust at high redshift is further supported by numerous semi-analytical and semi-numerical models \citep[e.g.][]{Valiante2011,Mancini2015,Mancini2016,Yung2018}.  In principle, detailed information regarding the dust content and distribution in high-redshift galaxies can be obtained from numerical simulations.  Examples of such simulations where the effects of dust at high redshift have been studied include Mare Nostrum \citep{Devriendt2010}, BlueTides \citep{Wilkins2018}, FIRE \citep{Ma2018}, FiBY \citep{Cullen2017}, and Renaissance \citep{Barrow2017}.  In particular, \cite{Wilkins2018} have shown that many galaxies at $z\gtrsim8$ are expected to be heavily dust obscured and the properties of such galaxies may be constrained in the future by ALMA.  However, numerous uncertainties remain regarding dust production and destruction (and hence dust mass), dust temperature, and dust composition/characteristics within simulations \citep{Dayal2018}.  Nevertheless, techniques for ``observing'' simulated galaxies to obtain, for instance, their SEDs are now commonplace and are providing detailed insight into the ISM properties of high redshift galaxies \citep[e.g][]{Zackrisson2013,Cen2014,Wilkins2016,Cullen2017,Barrow2017}.  For example, \cite{Behrens2018} used a numerical simulation to reproduce the dusty SED of a $z=8.38$ galaxy (A2744\_YD4, \citealt{Laporte2017}), at a slightly lower redshift than JD1. However, it remains to be determined whether the assembly histories of simulated galaxies can be reconciled with the strength of the Balmer break observed in JD1.

In this paper we examine the method of inferring cosmic dawn from the properties of several $z\simeq9$ galaxies, including JD1, using a cosmological, radiation hydrodynamics code with multifrequency radiative transfer. In addition to addressing the question of whether the assembly history of such sources inferred from observations are consistent with such models, we make predictions for further observables that may assist in breaking degeneracies between stellar ages and other parameters.

\begin{table*}
\centering
 \begin{tabular}{c|cccccccccccc} 
 \hline
ID & F435W & F606W & F775W & F814W & F850W & F105W & F125W & F140W & F160W & K$_s$ & 3.6$\mu$m & 4.5 $\mu$m \\
\hline
\hline
JD1 & $>$29.40 & $>$ 29.60 & - & $>$ 29.80 & - &$>$ 30.2 & 26.75 & 25.83 & 25.77 & $>$ 24.20 & 25.64 & 24.70 \\
  &   &   &   &   &   &  & $\pm$0.03 & $\pm$0.02 & $\pm$0.02 &  & $\pm$0.17 & $\pm$0.07 \\

GN-z10-3 &$>$28.70 & $>$28.90 & $>$28.53 & $>$ 28.39 & $>$ 28.39 & $>$28.15 & 27.99 & - & 26.74 & 26.58 & 27.42 & 26.48 \\
  &  &   &   &   &   &   & $\pm$0.38 &  & $\pm$0.12 & $\pm$3.33 & $\pm$0.59 & $\pm$0.24 \\
  
GN-z9-1 & $>$28.04 & $>$ 28.39 & $>$ 28.04 & $>$28.26 & $>$ 27.86 & $>$28.39 & 27.51 & 26.56 & 26.61 & 26.70 & 26.87 & 26.15 \\
  &   &   &   &   &   &   & $\pm$0.27 & $\pm$0.68 & $\pm$0.15 & $\pm$0.79 & $\pm$0.30 & $\pm$0.17 \\
  \hline
\end{tabular}
\caption{\label{photometry}Photometry of the three $z>$9 galaxies discussed in this study. Upper limits are given at 2$\sigma$. Values for GN-z10-3 and GN-z9-1 are from \citet{Oesch2014} and magnitudes for JD1 are from \citet{Hashimoto2018}. }
\end{table*}

This paper is organised as follows. In \S2 we select three $z\simeq9-9.5$ galaxies for this study and we briefly describe their observational data from which the SEDs and other properties are derived. In \S3 we summarise our numerical approach and illustrate its appropriateness for the topic at hand. We derive assembly histories and match the SEDs in \S4 and summarise our conclusions in \S5.

\section{Observational Data}
The key aspect of JD1 that makes it of particular interest in probing when cosmic dawn occurred is an IRAC channel 2 excess in its SED at 4.5 $\mu$m at the confirmed redshift $z=9.11$. This excess can only arise from starlight and hence provides a clear measurement of the strength of the Balmer break. The idea was first proposed by \cite{Zheng2012} and later \cite{Hoag2018,Hashimoto2018} demonstrated through careful modelling that such an excess in the SED cannot arise from any reasonable contamination by strong nebular emission lines such as [OIII] and H$\beta$. However, as discussed by \cite{Hashimoto2018}, the lensing magnification $\mu$, of JD1 is somewhat uncertain, formally ranging from $\mu\simeq10-80$. The adopted value naturally affects the absolute scale of the derived stellar mass assembly history and hence credibility of the star formation history in the context of structure formation models.

In addition to considering JD1, we have selected two further unlensed sources, GN-z10-3 and GN-z9-1 seen in the CANDELS survey, with similar IRAC channel 2 excesses from the compilation of $z\simeq9$ galaxies by \cite{Oesch2014}. Although not yet spectroscopically confirmed, both have photometric redshift likelihoods consistent with a similar Balmer break at about the same epoch. Table~\ref{photometry} summarises the photometric properties of the three selected sources and Figure 1 provides image stamps that illustrate the HST and Spitzer data. Naturally a key question is whether, without spectroscopic confirmation, we can follow the arguments for JD1 in ruling out an IRAC channel 2 excess from intense [O III] emission. Formally this requires both sources to have redshifts beyond $z\simeq9.0$. In Figure 2 we show the photometric likelihood distribution for both sources alongside that for JD1 for which spectroscopic confirmation was finally provided by \cite{Hashimoto2018}. As can be seen, it seems reasonably likely that the IRAC channel 2 excess in GN-z10-3, and to a lesser extent, GN-z9-1 also arise from a Balmer break.

In analysing the star formation histories and dust content of all 3 sources, we seek to go beyond understanding whether they share a common origin but also to examine whether their derived properties are consistent with numerical simulations of early galaxies. In this respect however, it is important to realise that these three sources have been carefully selected from the population of HST-located sources in both the deep fields (the Ultra Deep Field and CANDELS surveys) and the lensed surveys (principally the Hubble Frontier Field campaign). As they are chosen purely on the basis of their likely redshift and Balmer break, they may not be representative of the overall population of luminous sources at $z\simeq9$. Nevertheless, as they provide the most useful early constraints on cosmic dawn, it is important to understand whether their properties can be reproduced by cosmological simulations.

\begin{figure}
\centering
\includegraphics[width=10cm,trim={0.7cm 1.6cm 0cm 1.7cm},clip]{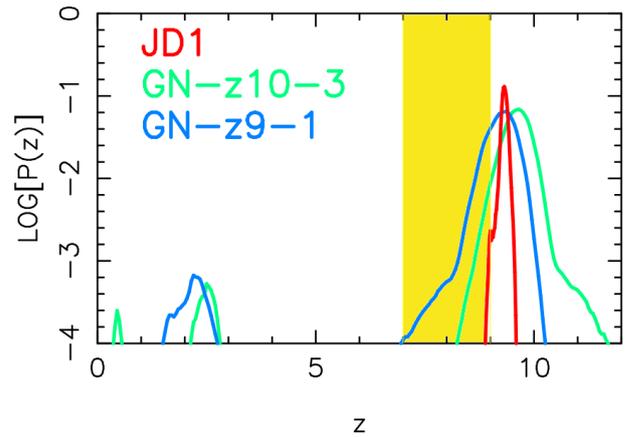}
\caption{Redshift probability distribution for the 3 candidates discussed in this study, namely JD1 (red), GN-z9-1 (blue) and GN-z10-3 (green). The yellow box displays the redshift window where the colours observed between IRAC channel 1 and 2 could be explained by the contamination of the 4.5$\mu$m photometry by strong [OIII]+H$\beta$ emission lines. The P(z) of our 3 objects only enter marginally into this redshift window, in favour of a Balmer break explanation for the red IRAC colours observed.}
\end{figure}

\begin{figure*}
\centerline{\includegraphics[scale=1]{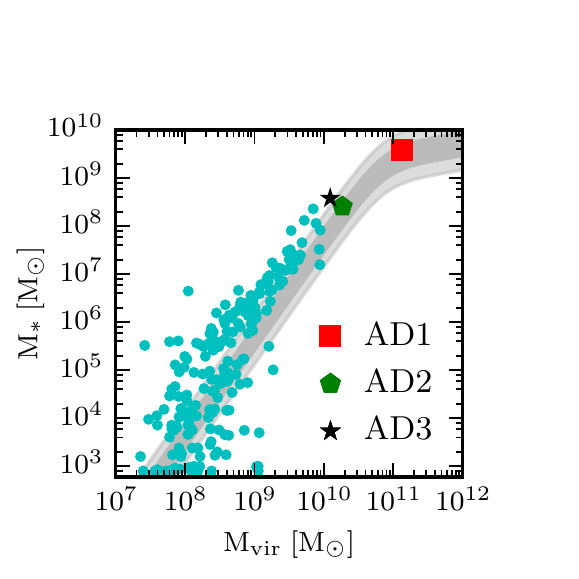}\includegraphics[scale=1]{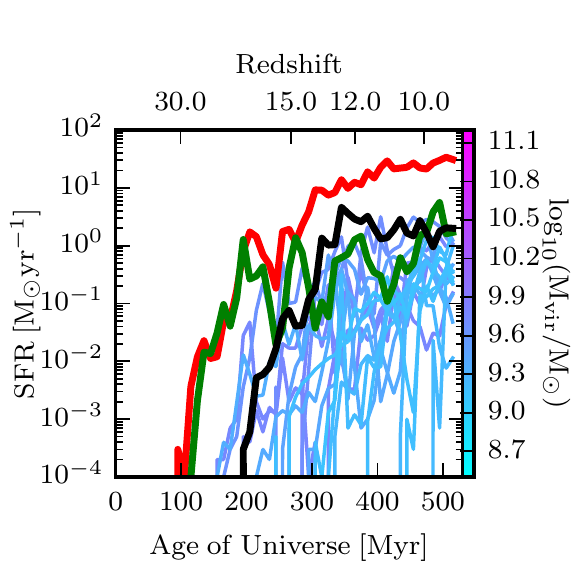}\includegraphics[scale=1]{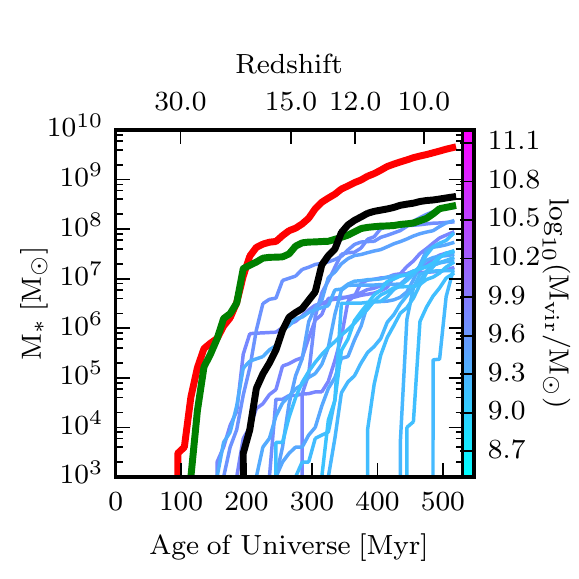}}
\caption{(Left) Stellar mass-halo mass relation for all uncontaminated central systems within the zoom-in region that are resolved by at least 300 DM particles at $z=9.2$.  The grey shaded region shows the $1\sigma$ and $2\sigma$ scatter on the extrapolated predictions from abundance matching from \protect\citep{Behroozi2013}.  The red square, green pentagon, and black star represent AD1, AD2, and AD3, respectively, while the cyan points represent other resolved haloes in the simulation that host stars.   (Centre)  SFR as a function of the age of the Universe for the same set of galaxies.  Red, green, and black represent AD1, AD2, and AD3, while the other lines are coloured based on their virial mass. (Right) Cumulative stellar mass as a function of the age of the Universe.}
\label{stars}
\end{figure*}

\section{Numerical Simulations}
\label{numsim}

To compare with the observed galaxies, we perform a cosmological gravitational radiation-hydrodynamics zoom-in simulation of a massive Lyman-break galaxy (LBG) at $z>6$, using the publicly available, adaptive mesh refinement (AMR) code {\small RAMSES-RT} \citep{Teyssier2002,Rosdahl2013,Rosdahl2015}.  Initial conditions for a low-resolution ($256^3$) dark matter-only simulation were generated in a box with side length 50~comoving~Mpc with {\small MUSIC} \citep{Hahn2011}.  A dark matter halo with mass ${\rm M_{\rm vir}=10^{11.8}M_{\odot}}$ was selected at $z=6$ for re-simulation.  New initial conditions were iteratively generated around the Lagrange region of the system of interest using a convex-hull until the high-resolution dark matter only simulation resulted in the halo being uncontaminated by low resolution dark matter particles out to $2R_{\rm vir}$ at $z=6$.  In the final set of optimised initial conditions, the set of high resolution particles were generated to have an effective resolution of $4096^3$ particles, corresponding to a DM particle mass of $4\times10^4{\rm M_{\odot}}h^{-1}$.  The cosmology of the initial conditions was set so that $h=0.6731$, $\Omega_{\rm m}=0.315$, $\Omega_{\rm b}=0.049$ , $\Omega_{\Lambda}=0.685$, $\sigma_8=0.829$, and $n_s=0.9655$, consistent with \cite{Planck2016}.  All gas in the box is assumed to be initially neutral and composed of 76\% H and 24\% He by mass. 

We use the version of {\small RAMSES} presented in \cite{Katz2017,Kimm2017} to model the gravity, hydrodynamics, radiation transfer, and non-equilibrium chemistry in the cosmological box.  The radiation is evolved using a moment method using the M1 closure for the Eddington tensor \cite{Levermore1984}.  The radiation is split into eight frequency bins: infrared, optical, Habing, Lyman-Werner, H-ionizing, H$_2$ ionising, He-Ionising, and HeII-ionising, as listed in Table~2 of \cite{Kimm2017}.  The radiation is coupled to the gas by photoionisation, photoheating, and radiation pressure (both UV and IR).  Details of the thermochemistry model for H and He can be found in \citet{Rosdahl2013}, details regarding the H$_2$ implementation, cooling, and coupling to radiation can be found in \citet{Katz2017,Kimm2017}, while the radiation pressure description is presented in \citet{Rosdahl2015}.  In addition to gas cooling from primordial species, we also employ temperature and density metal line cooling tables that have been computed with {\small CLOUDY} \citep{Ferland1998} for $T>10^4$K.  At lower temperatures, we employ the cooling rates from \citet{Rosen1995}.  In order to model the radiation pressure on dust, we assume a mean dust opacity of 10cm$^2$g$^{-1}$ in the IR radiation bin and 10$^3$cm$^2$g$^{-1}$ in all other radiation bins, consistent with \citet{Rosdahl2015b}.  The dust mass for each cell is computed using the metallicity dependent dust-to-metal ratios from \cite{Remy2014} as implemented into {\small RAMSES} by \cite{Kimm2018}.  Furthermore, we assume that the dust has been destroyed in all cells with $T>10^5$K.  

Because we employ an explicit solver to model the advection of radiation, the simulation time step is limited by the RT-courant condition.  For this reason, we reduce the speed of light in the simulation to $c_{\rm sim}=0.01c$ on all AMR levels to reduce the constraint on the global time step.  Since we do not aim to model the global reionization process and only focus on a very over-dense region of the Universe, this approximation is justified (see \citealt{Katz2017}).  Even with this reduced speed of light, at early epochs, the time step in the simulation is still dominated by the RT-courant condition.  Hence we subcycle the radiation up to 500 times on each level thereby reducing the number of hydrodynamic time steps that need to be computed.  The method for RT subcycling can be found in \cite{Commercon2014,Rosdahl2018}.  

The only sources of radiation in the simulation are star particles.  Star formation is modelled using a Schmidt law \citep{Schmidt1959} where the star formation rate density is related to the gas density divided by the gas free-fall time, modulated by an efficiency parameter.  Stars are only allowed to form inside the zoom-in region and on the maximum AMR refinement level.  We search cells with $\rho_{\rm gas}>100{\rm cm^{-3}}$ to determine whether their thermo-turbulent Jeans length is unresolved (see Equation~8 of \citealt{Kimm2017}).  If a cell satisfies these conditions, we then compute the efficiency of star formation based on the thermodynamic properties of the cell (see Equation~2 of \citealt{Kimm2017}).  With this efficiency, the number of newly formed star particles is drawn from a Poisson distribution with a minimum stellar mass of 1000M$_{\odot}$.

\begin{table*}
\centering
\begin{tabular}{@{}lccccccccc@{}}
\hline
Redshift & $z=12.0$ & $z=11.5$ & $z=11.0$ & $z=10.5$ & $z=10.0$ & $z=9.8$ & $z=9.6$ & $z=9.4$ & $z=9.2$\\ 
\hline
\hline
{\bf AD1} & & & & & & & & &\\ 
Halo Mass & 10.38 & 10.46 & 10.54 & 10.62 & 10.83 & 10.91 & 10.98 & 11.02 & 11.06\\ 
Stellar Mass & 8.76 & 8.89 & 9.04 & 9.13 & 9.37 & 9.41 & 9.49 & 9.54 & 9.58\\ 
Gas Mass & 9.51 & 9.60 & 9.68 & 9.75 & 10.02 & 10.08 & 10.15 & 10.19 & 10.22\\ 
SFR & 6.21 & 13.99 & 18.51 & 8.23 & 16.51 & 20.37 & 25.99 & 29.35 & 24.10\\ 
Metallicity & 1.87e-3 & 1.96e-3 & 2.15e-3 & 2.71e-3 & 2.57e-3 & 2.49e-3 & 2.45e-3 & 2.51e-3 & 2.61e-3\\ 
H$_2$ Mass & 8.26 & 8.35 & 8.41 & 8.67 & 8.90 & 8.92 & 8.96 & 8.99 & 9.04\\ 
$r_{\rm e}$ & 0.16/1.43 & 2.12/2.19 & 0.10/0.71 & 0.12/0.66 & 0.13/0.68 & 0.12/0.72 & 0.10/0.75 & 0.10/0.66 & 0.73/1.02 \\
$A_{1500}$ & 1.78/1.98 & 1.02/1.08 & 3.22/3.31 & 2.86/3.01 & 2.46/2.46 & 2.67/2.72 & 2.56/2.54 & 3.27/3.22 & 2.83/2.85\\ 
\hline
{\bf AD2} & & & & & & & & &\\ 
Halo Mass & 9.96 & 10.01 & 10.08 & 10.13 & 10.16 & 10.18 & 10.20 & 10.21 & 10.22\\ 
Stellar Mass & 7.83 & 7.94 & 7.96 & 8.01 & 8.10 & 8.16 & 8.34 & 8.37 & 8.41\\ 
Gas Mass & 8.99 & 9.07 & 9.21 & 9.31 & 9.38 & 9.43 & 9.45 & 9.50 & 9.52\\ 
SFR & 0.78 & 0.47 & 0.24 & 0.53 & 1.28 & 1.39 & 5.34 & 1.28 & 1.96\\ 
Metallicity & 5.98e-4 & 7.46e-4 & 6.64e-4 & 5.75e-4 & 5.75e-4 & 5.95e-4 & 7.05e-4 & 8.74e-4 & 9.81e-4\\ 
H$_2$ Mass & 7.28 & 7.08 & 7.18 & 7.55 & 7.42 & 8.01 & 7.89 & 7.87 & 8.03\\ 
$r_{\rm e}$ & 0.18/0.12 & 0.17/0.16 & 0.26/0.26 & 0.24/0.25 & 0.08/0.22 &  0.48/2.42 & 2.74/2.76 & 2.61/2.74 & 0.18/0.33 \\
$A_{1500}$ & 0.40/0.78 & 0.18/0.43 & 0.11/0.10 & 0.13/0.12 & 0.50/0.63 & 0.30/0.41 & 0.54/0.52 & 0.44/0.41 & 1.21/1.41 \\
\hline
{\bf AD3} & & & & & & & & &\\ 
Halo Mass & 9.71 & 9.77 & 9.82 & 9.87 & 9.92 & 9.93 & 9.95 & 9.98 & 10.03\\ 
Stellar Mass & 8.11 & 8.25 & 8.31 & 8.40 & 8.48 & 8.49 & 8.51 & 8.54 & 8.58\\ 
Gas Mass & 8.91 & 8.95 & 9.00 & 9.05 & 9.11 & 9.13 & 9.15 & 9.20 & 9.27\\ 
SFR & 2.91 & 2.94 & 1.20 & 2.43 & 2.18 & 1.12 & 1.42 & 1.66 & 2.06\\ 
Metallicity & 1.28e-3 & 2.12e-3 & 2.69e-3 & 2.83e-3 & 3.00e-3 & 3.09e-3 & 3.09e-3 & 2.93e-03 & 2.59e-3\\ 
H$_2$ Mass & 7.41 & 7.63 & 8.09 & 8.08 & 8.10 & 8.30 & 8.31 & 8.32 & 8.30\\ 
$r_{\rm e}$ & 0.07/0.18 &0.07/0.15 &0.10/0.24 &0.07/0.29 &0.08/0.36 &0.08/0.40 &0.08/0.41 &0.10/0.46 & 0.08/0.46 \\
$A_{1500}$ & 2.30/2.24 & 2.94/2.29 & 3.05/3.15 & 2.80/3.07 & 2.80/3.01 & 2.01/2.60 & 2.69/3.10 & 2.91/3.08 & 2.55/2.83 \\
\hline
\end{tabular}
\caption{Simulated galaxy properties as a function of redshift.  All masses are quoted in $\log_{10}({\rm M/M_{\odot}})$ and radii are listed in kpc.  Metallicities represent the mean mass-weighted gas-phase metallicity, while the SFRs are in units of ${\rm M_{\odot}/yr}$ averaged over the previous 10Myr.  $r_{\rm e}$ represents the effective radius of the galaxy.  The two values show the results from the face-on configuration in the case without dust, and with dust, respectively.  $A_{1500}$ represents the number of magnitudes of extinction at 1500$\AA$.  The two numbers represent the face-on and side-on configurations of the galaxy, respectively. The morphology can change rapidly due to disruption from stellar feedback and mergers, hence also changing the direction of the principle angular momentum axis.}
\label{hprops}
\end{table*}

\begin{figure*}
\centerline{\includegraphics[scale=1]{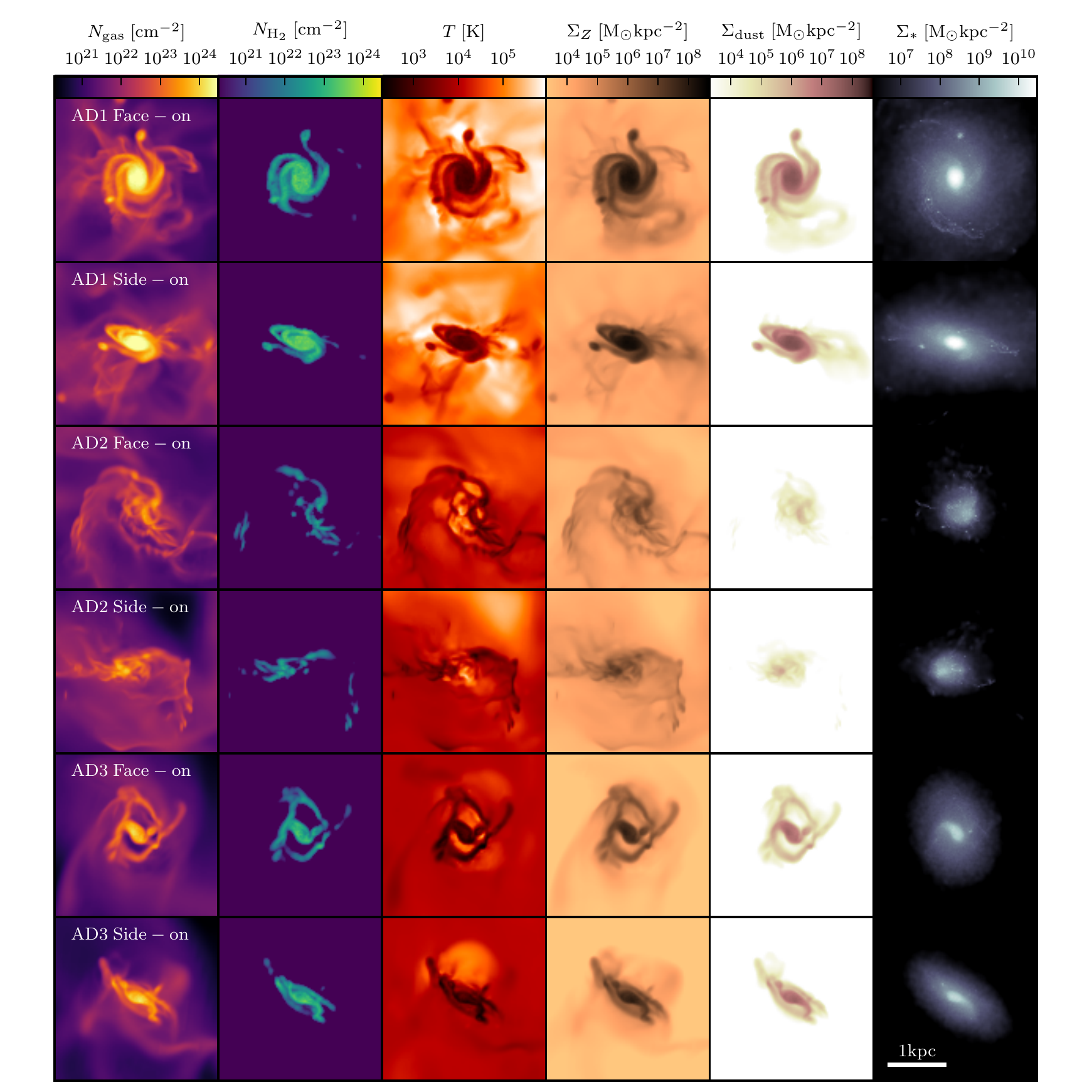}}
\caption{Maps of total gas column density, H$_2$ column density, density weighted temperature, metal surface mass density, dust surface mass density, and stellar mass surface density for our three systems, AD1, AD2, and AD3 in both face-on and side-on views at $z=9.6$.}
\label{allprops}
\end{figure*}

Once a star is formed, it injects radiation into its host cell.  The spectra of each star particle is interpolated based on age and metallicity using the {\small BPASSv2} model \citep{Stanway2016,Eldridge2008}, assuming a maximum stellar mass of 300M$_{\odot}$.  The total luminosity of each spectra is scaled with the total mass of the star particle.  Stars also impact the gas via supernova (SN) feedback.  For each SN, $10^{51}$ergs is injected in the form of momentum and these occur during the first 50Myr of the lifetime of the star particle by sampling a delay time distribution.  The model we use for momentum injection is presented in \cite{Kimm2015,Kimm2017,Rosdahl2018} and it is designed to capture the right amount of momentum at the end of the snowplow phase which is dependent on the resolution and properties of the cell that hosts the star particle.  We assume a mass fraction of 0.2 is recycled back into the gas for each star particle.  Metals are also returned with this gas assuming a metallicity of 0.075.  In order to calibrate the feedback to reproduce the high redshift stellar mass-halo mass relation, we follow \cite{Rosdahl2018} and boost the total number of SN for each star particle by a factor of four.  In the left panel of Figure~\ref{stars}, we show the stellar mass-halo mass relation for all uncontaminated galaxies at $z=9.2$ compared with the predictions from abundance matching \citep{Behroozi2013} and there is a strong agreement between the two due to the calibrated feedback.  In the centre and right panels, we show the SFR histories of the 20 most massive systems and the stellar mass assembly histories of the same systems.  We have highlighted the three most massive systems, AD1, AD2, and AD3 in red, green, and black, respectively, on all three plots.  These are the three systems we will focus on in the majority of our analysis and it is clear that even among these three objects, there is a diversity in star formation and mass assembly histories that will affect the shapes of the SEDs.

As the simulation evolves, we allow the cells to adaptively refine to obtain higher spatial and mass resolution in the regions of interest.  We allow for up to 19 total levels of refinement in order to maintain a roughly constant maximum physical resolution of $\sim13.6$pc throughout the simulation.  Cells are refined in a quasi-Lagrangian fashion when cells have DM masses of eight times the particle mass or gas masses that are eight times $\frac{\Omega_{\rm b}}{\Omega_{\rm DM}}m_{\rm DM}$.  Furthermore, we refine cells when the Jeans length is not resolved by at least four cell lengths.  When a cell is refined, it is split into eight children cells.

After the simulation is completed, we run a halo finder on the snapshots in order to isolate the halo locations and characterise their properties.  We use the AMIGA halo finder \citep[AHF,][]{Gill2004,Knollmann2009}, and set the virial radius to be such that the mean density of the halo would allow for collapse assuming a spherical over-density against an expanding background for our cosmology at each redshift.  At the redshifts we sample, this value corresponds to $\Delta\sim200$.  We only consider haloes that are not contaminated by any low resolution dark matter particles and that have a DM halo made up of at least 300 DM particles.  This sets a lower limit to the halo mass we consider of $1.2\times10^7{\rm M_{\odot}}h^{-1}$ which is well below the atomic cooling threshold.

\begin{figure*}
\centerline{\includegraphics[scale=1]{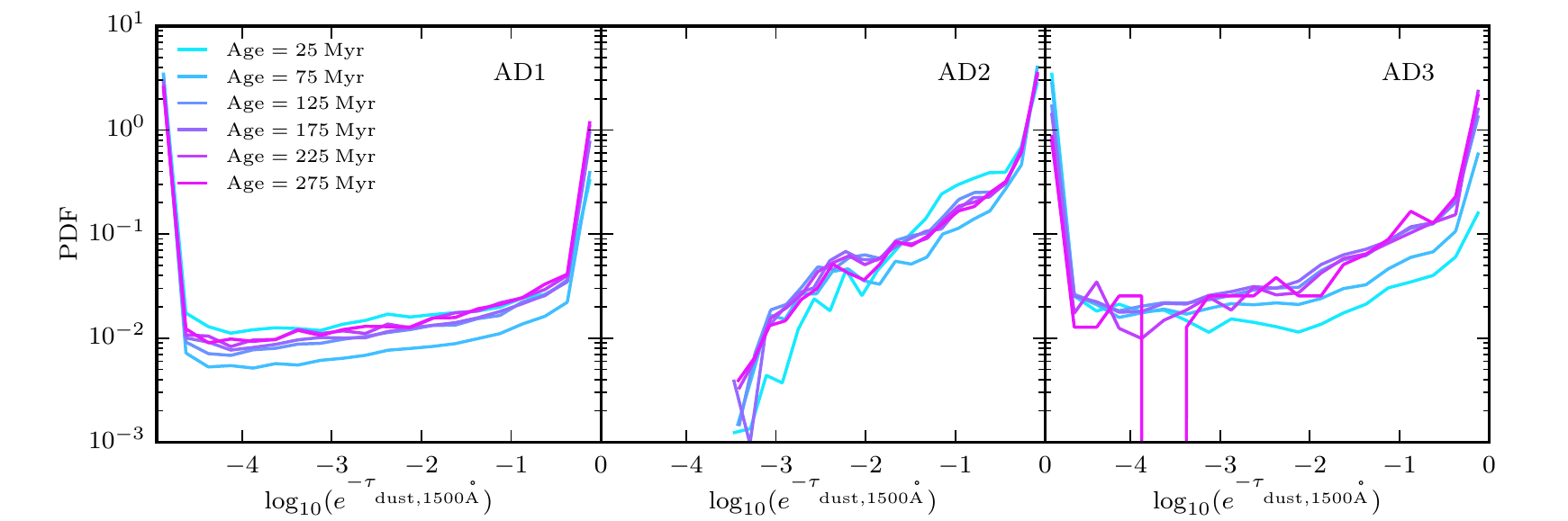}}
\caption{Mass-weighted probability distribution functions (PDFs) of the logarithm of the transmission at 1500\AA~($\log_{10}(e^{-\tau_{\rm dust,1500\AA}})$) for stars binned by stellar age in each of our three galaxies, AD1 (left), AD2 (centre), and AD3 (right) at $z=9.6$.  The legend indicates the mean age of the stellar population and the width of the bins is 50Myr.  High values of $\log_{10}(e^{-\tau_{\rm dust,1500\AA}})$ indicate that most of the emission escapes the galaxy unattenuated while very low values indicate that much of the emission is absorbed.  For AD1 and AD3, almost all of the very young star particles with age~$<50$Myr have very low transmission demonstrating a strong attenuation by dust while many of the older star particles have high transmission after SN feedback has blown away the dense surrounding gas.  In contrast, AD2 has very high transmission for all stellar ages.  The morphology of AD2 is considerably different than AD1 and AD3 (see Figure~\ref{allprops}).}
\label{tau_dust}
\end{figure*}

To compute the SEDs of each of the simulated galaxies, we use the age, metallicity, and mass of each star particle to compute the shape and normalisation of the intrinsic SED.  Here we have used the {\small BPASSv2} model \citep{Stanway2016,Eldridge2008}, assuming a maximum stellar mass of 300M$_{\odot}$; however, our results are similar when using the BC03 SED as used in \cite{Hashimoto2018}.  We then compute the optical depth of dust to each star particle (following the same dust-to-gas ratio as used in the simulation when modelling IR radiation pressure) for a given viewing angle (either either face-on or side-on) using the $R=3.1$ dust grain model of \cite{Weingartner2001} and attenuate the SED.  Summing over all star particles gives the total stellar continuum SED of the galaxy.  In addition to the stars, we add the contribution from the nebular continuum which has been shown to be important for the total SED \citep[e.g.][]{Wilkins2016}. Details of that calculation are described in Appendix~\ref{methods} and this contribution is also dust attenuated in the same way as the star particles.  Note that at the redshift of interest, the Balmer break of JD1 is likely unaffected by nebular lines and hence we only focus on the continuum.  Our modelling neglects scattering by dust which is quite computationally expensive as it requires post-processing each galaxy with radiation transfer.  The effects of neglecting scattering are described in \cite{Kaviraj2017} where they compare the {\small SUNSET} code (which uses a very similar method to that in our work) with the full dust radiation-transfer code {\small SUNRISE} \citep{Jonsson2006,Jonsson2010}.

\begin{figure}
\centerline{\includegraphics[scale=1]{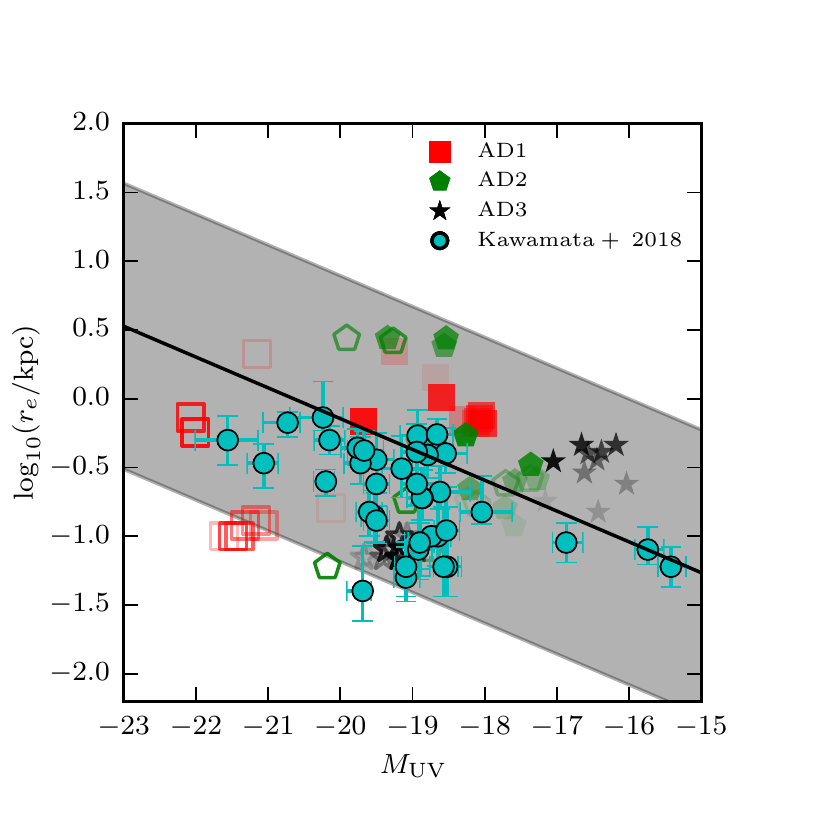}}
\caption{UV Magnitude-size relation for galaxies in our simulations compared with the $z\sim9$ observations of \protect\cite{Kawamata2018}.  Cyan points represent observations while the red squares, green pentagons, and black stars represent AD1, AD2, and AD3, respectively at different redshifts.  More translucent points are high redshift and darker points are lower redshifts.  The open symbols represents the effective stellar radius when we do not account for dust attenuation while filled points represent the same quantity when we include dust attenuation.  The black line and grey band represent the completeness corrected relation from \protect\cite{Kawamata2018} and 1$\sigma$ uncertainty.}
\label{slz9}
\end{figure}

\section{Results}
Determining whether the simulations described in this paper can reproduce the strong Balmer breaks observed at high redshifts is the main goal of this work.  Although the simulation contains thousands of galaxies, we focus on the three most massive systems, AD1, AD2, and AD3, which have halo masses of $10^{11.1}$, $10^{10.2}$, and $10^{10.1}$M$_{\odot}$ at $z=9$, respectively.  A full list of properties for each galaxy as a function of redshift in the range $9.2\leq z\leq 12$ can be found in Table~\ref{hprops}.  Throughout the range in redshifts, the galaxies exhibit a diverse set of metallicities, star formation rates, and stellar masses that may be representative of the high-$z$ galaxies that have already been observed.

\subsection{Structural Properties}
In Figure~\ref{allprops}, we show the gas column density, H$_2$ column density, density weighted temperature, metal surface mass density, dust surface mass density, and stellar surface mass density for each of the three galaxies at $z=9.6$ for two viewing angles with the primary angular momentum axis either oriented into the page (face-on) or in the plane of the page (edge-on).  At this redshift, AD1 exhibits a well defined cold, gaseous disk with strong features indicating ordered rotation.  The molecular content of the galaxy is confined to the disk and the temperature can reach $\gg 10^5$K surrounding the cold , $\ll 10^3$K, central region.  The H$_2$ primarily follows the metals and dust, which are also very prominent in the disk, as the formation rate of H$_2$ on dust is much more efficient than via the primordial H$^-$ channel.  In the bottom right of the face-on metal and dust map for AD1, a plume of metals and dust can be seen moving away from the galaxy due to SN feedback driving enriched material away from the system.  Note that the dust map is not a simple rescaling of the metallicity map as the dust-to-metal ratio scales with metallicity and we have also assumed a constant dust sublimation temperature of $10^5$K, hence there being almost no dust outside of the cold, central disk.  Inside the central regions of the galaxy, the dust surface mass density can reach $>10^8{\rm M_{\odot}kpc^{-2}}$.  The last column of Figure~\ref{allprops} shows the distribution of stellar mass in the galaxy.  There is a bright central region with a surrounding halo and two stellar arms, curling over the top and under the bottom of the central region in the face-on image of AD1.  Additionally, there is a small star-forming clump at a projected distance of $\sim750$pc above the central cluster.  Comparing the dust map with the stellar map, it is evident that the dust is preferentially located in the same regions as the stellar mass and this will indeed be reflected in the SED of the galaxy.

While AD1 appears to be a well-ordered, rotating disk, AD2, in contrast, is significantly more diffuse and does not contain any ordered structure.  The stellar mass is distributed more evenly throughout the central region, without any strong, central bulge and the surface mass density of dust is significantly reduced compared to AD1.  

Despite having very similar halo mass to AD2, AD3 is more reminiscent of AD1, and has a dense central region with signs of rotation.  The column density of dust in AD3 is significantly higher than that of AD2, and similar to AD1. The strong features in the stellar mass distribution of AD3 are also reflected in the dust map; hence, a significant amount of attenuation is expected for AD1 and AD3.

\subsection{The Spatial Distribution of Dust}
Understanding the spatial distribution of dust is key for predicting the amount of attenuation in high-redshift galaxies.  If the spatial distribution is biased towards a specific stellar population, this could affect the interpretation of the Balmer break.  We have measured the mass weighted probability distribution function of the dust attenuation, $\log_{10}(e^{-\tau_{\rm dust,1500\AA}})$, for different populations of stars, binned by stellar age, for one viewing angle (down the $x$-axis of the simulated box) at $z=9.6$ for each of the three galaxies (see Figure~\ref{tau_dust} but note that the exact curves will change based of viewing angle and the affect of viewing angle on the SED is discussed later).  For AD1, and AD3, the strongest peak in the PDF occurs at $\log_{10}(e^{-\tau_{\rm dust,1500\AA}})=-5$ where nearly all of the 1500\AA~luminosity is absorbed.  The PDF has a second smaller peak at $\log_{10}(e^{-\tau_{\rm dust,1500\AA}})=0$ where we expect complete transparency at this wavelength.  Interestingly, in the right panel of Figure~\ref{tau_dust}, the attenuation PDFs for AD3 for different stellar age bins deviate from one another when $\log_{10}(e^{-\tau_{\rm dust,1500\AA}})\rightarrow0$ such that the young stars have a much lower peak in the PDF at high transmission.  The probability that the line of sight is completely optically thin for a set of old stars is $>10$ times more likely compared to the youngest stars in AD3.  This difference is not seen as strongly for AD1 and thus AD3 is our best example of a galaxy where the UV radiation from the youngest stars is most preferentially absorbed by dust. Hence AD3 is a very good example of a galaxy where the strength of the Balmer break may provide misleading information about the {\bf total} stellar population and assembly history.

The dust attenuation PDF for AD2 has a completely different shape than that of AD1 or AD3.  The strongest peak occurs at $\log_{10}(e^{-\tau_{\rm dust,1500\AA}})=0$ where we expect complete transmission for UV photons.  The PDF falls off very steeply towards stronger attenuation factors.  The diffuse morphology of AD2 is the culprit for this effect. Thus, despite having a nearly identical stellar and halo mass to AD3, these two galaxies have different UV magnitudes.

In Table~\ref{hprops}, we list the amount of attenuation at a wavelength of 1500$\AA$ ($A_{1500}$) for each galaxy as a function of redshift for two different configurations, either with the principal angular momentum axis perpendicular (face-on) or parallel (side-on).  Depending on the instantaneous state of the galaxy and the orientation, the amount of attenuation can change rapidly.  For example, the amount of attenuation for AD2 viewed from side-on changes by more than a factor of 5 between $z=10.5$ and $z=10.0$.  Comparing with other simulations, \cite{Ma2018} find that some of their brighter galaxies can exhibit more than three magnitudes of attenuation at a fixed UV magnitude, consistent with AD1 and AD3.  \cite{Wilkins2018} also find that $A_{1500}$ can scatter to values $>3$.  However, each of these simulations was run with different subgrid models, different dust models, and at different spatial resolutions.  We only probe three systems in our current work so a larger sample will be required for a systematic comparison with other simulations.

The spatial distribution of the dust has a strong impact on the perceived size of a galaxy.  We demonstrate this in Figure~\ref{slz9} where we show the stellar effective radius ($r_{\rm e}$) versus the UV magnitude of our simulated galaxies viewed from face-on compared with the $z\sim9$ observations of \cite{Kawamata2018}.  Open symbols represent the intrinsic distributions while filled symbols show the results after accounting for dust attenuation.  The intrinsic and dust attenuated values of $r_{\rm e}$ are listed in Table~\ref{hprops}.  Without dust attenuation,  $r_{\rm e}$ tends to scatter to small values, sometimes $<100$pc and this is especially true for AD1 and AD3.  In contrast, with dust attenuation the effective radius significantly increases and all of our systems are in very good agreement with the completeness corrected relation from \cite{Kawamata2018}.

\subsection{The Impact of Dust on the SED}
\begin{figure*}
\centerline{\includegraphics[scale=1]{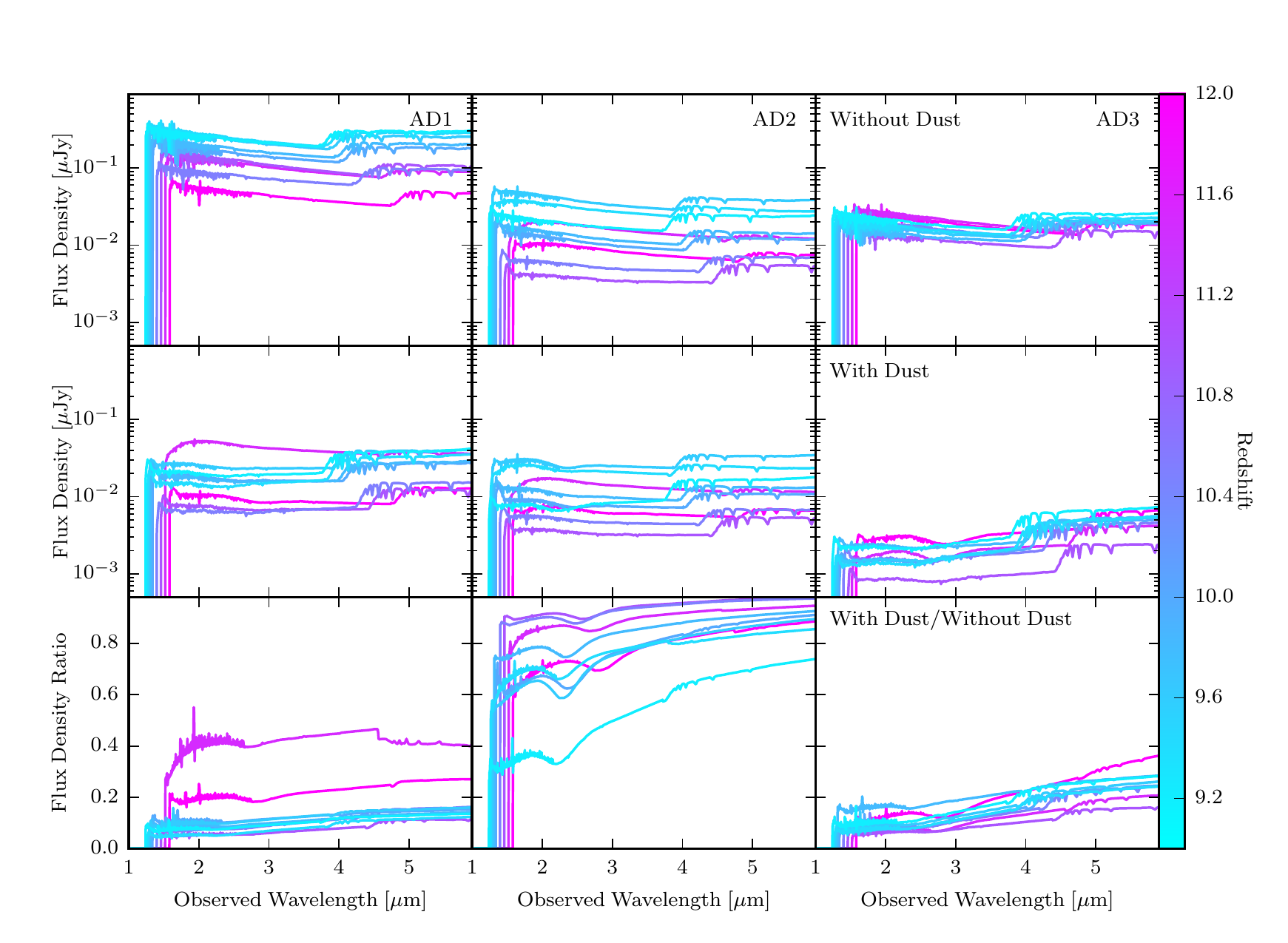}}
\caption{SEDs as a function of redshift for our three galaxies.  The top row shows the SEDs without any dust attenuation while the middle row includes dust.  The bottom row shows the ratio of the dust attenuated SED to the intrinsic SED to demonstrate how much absorption occurs at each wavelength. Balmer breaks are very unlikely in our simulated galaxies without dust.}
\label{zsed}
\end{figure*}

To demonstrate the impact of different dust distributions on the galaxy SEDs, in Figure~\ref{zsed} we plot the intrinsic SEDs (top row), the dust attenuated SED (middle row), and the ratio between the two (bottom row) for each of the three galaxies as a function of redshift in the range $12\geq z\geq 9.2$, for a face-on viewing angle.  All systems have been placed at the luminosity distance corresponding to the redshift of the simulation snapshot where we study the object and we have compared observed flux density in $\mu$Jy to observed wavelength in $\mu$m.  The different morphologies of the galaxies clearly lead to significantly different amounts of attenuation.  The key point we aim to emphasise with Figure~\ref{zsed} is that since the dust is found preferentially around young stars in both AD1 and AD3, the strength of the Balmer break is clearly enhanced between the intrinsic and dust attenuated SEDs (top and middle rows) because of the enhanced contribution from older stars to the total SED.

In the following subsections, we will compare the SEDs of our three simulated galaxies to the observations of JD1, GN-z10-3, and GN-z9-1 to determine whether the distribution of dust in the simulations is enough to reproduce the observed photometry with the realistic star formation and mass assembly histories shown in Figure~\ref{stars}.

\subsection{JD1}
\begin{figure*}
\centerline{\includegraphics[scale=1]{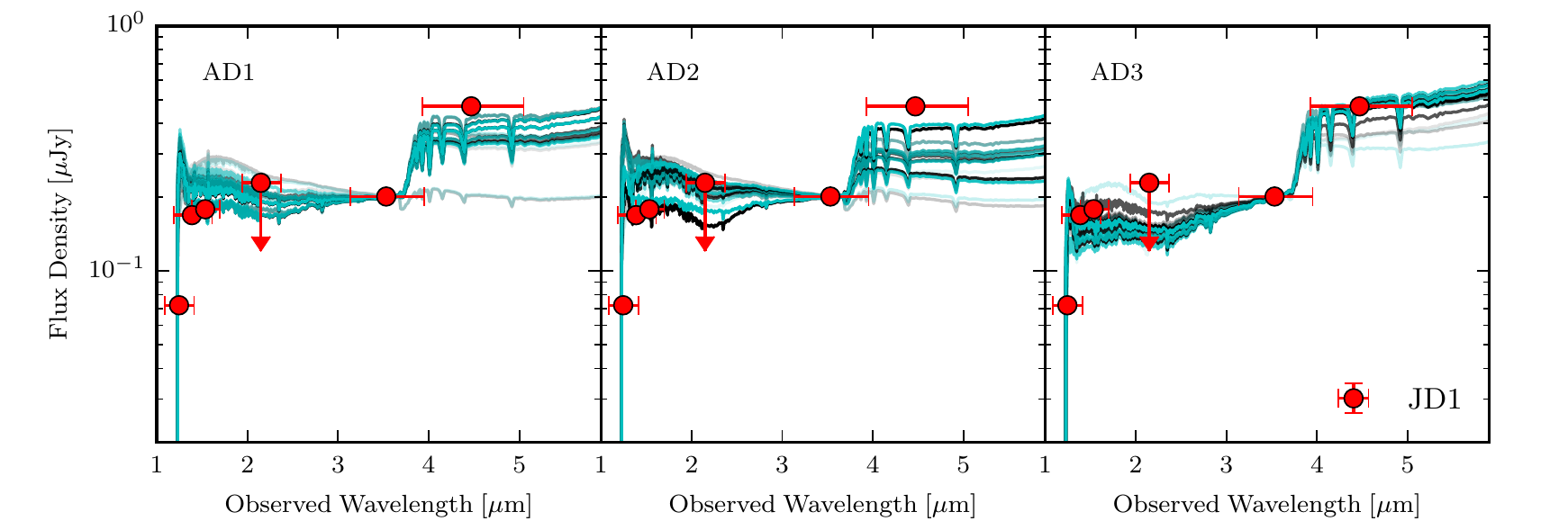}}
\caption{Comparison of the continuum SEDs of AD1, AD2, and AD3 from different redshifts with the photometry of JD1 (shown as the red points).  Note that the $K$-Band value for JD1 represents a 2$\sigma$ upper limit.  We assume that all SEDs are at $z=9.1$ to compare with the JD1 photometry.  Lighter lines indicate galaxies from higher redshift ($z\sim12$) snapshots while darker lines indicate lower redshifts down to $z=9.2$.  The black, and cyan lines depict viewing angles of either face-on or side-on, respectively.}
\label{JD1a}
\end{figure*}

We first compare with JD1, which is the only galaxy in our small sample that has been spectroscopically confirmed, and because of Frontier Fields data quality, has the most precise photometry.  In Figure~\ref{JD1a}, we compare the SEDs of AD1, AD2, and AD3 at all redshifts and two different viewing angles with the photometry of JD1.  We have shifted each simulated SED to $z=9.1$, and renormalised the SEDs to match the flux density in the IRAC channel~1 to compare whether the strength of the Balmer break in JD1 can be reproduced at any redshift by our simulated galaxies.  The renormalisation is akin to changing the stellar mass while keeping the same star formation history shape.  This can also be interpreted as requiring magnification if the flux density needs to be increased to match the photometry of JD1.

At high redshifts, shown as the lighter lines in Figure~\ref{JD1a}, the SEDs tend to be very blue for AD1 and AD3 with very weak Balmer breaks that are completely inconsistent with that of JD1.  As redshift decreases, the first stellar populations age, and young stars preferentially form in the dustiest regions of the galaxy, enhancing the strength of the Balmer break. Of the three galaxies, AD3, our lowest mass system, is found to have an SED that can reproduce the strength of the Balmer break in JD1, demonstrating that such a feature can be naturally reproduced by cosmological simulations.

\begin{figure*}
\centerline{\includegraphics[scale=1]{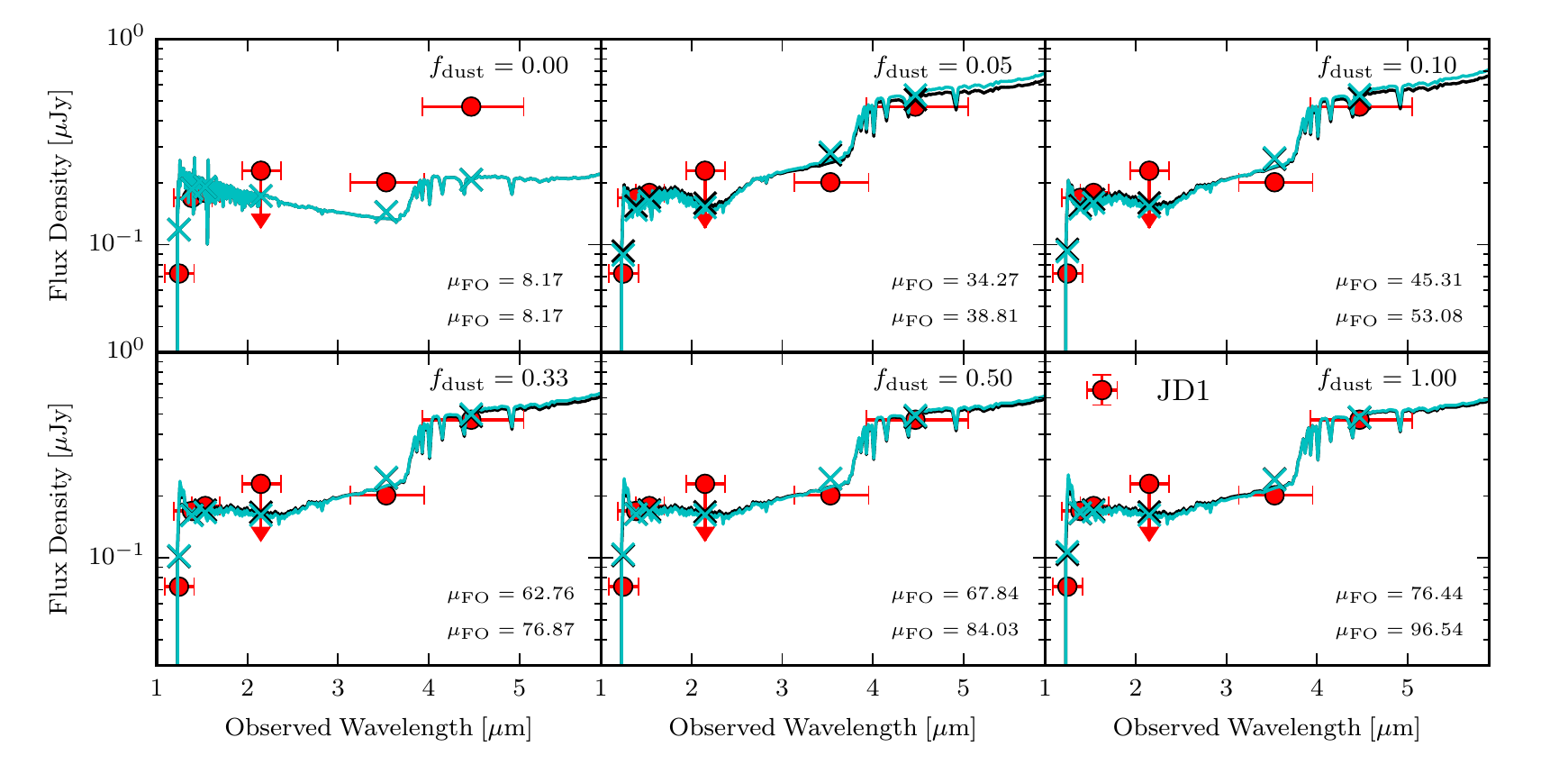}}
\caption{Comparison of the AD3 SED at $z=9.2$ with the photometry of JD1 (shown as red points).  The top left panel shows the unattenuated AD3 SED while the other five panels show the SED with different normalisations of gas-to-dust ratio, with $f_{\rm dust}$ indicated on the plot.  The bottom right panel is our fiducial model.  All models have been renormalised to minimise the $\chi^2$ of the model SED to the JD1 photometry.  The two viewing angles, face-on and side-on, are shown in black and cyan, respectively.  The magnification needed to re-normalise the SED for each viewing angle is listed in the plot as $\mu_{\rm FO}$ and $\mu_{\rm SO}$.  The crosses indicate the flux density in each photometric band while the lines show the continuum SED.  The strength of the Balmer break in JD1 can nearly be reproduced by AD3 when realistic dust masses and distributions are assumed.}
\label{JD1}
\end{figure*}

This is further explored in Figure~\ref{JD1} where we show the $z=9.2$ SED of AD3 (which has one of the largest Balmer breaks at a very similar redshift) compared with the photometry of JD1, where we have now explicitly tried to minimise the $\chi^2$ of the fit of the AD3 SED to the JD1 photometry by varying the amount of magnification needed to fit the normalisation of the SED.  Because of the large uncertainties on stellar yields, dust production mechanisms, dust-to-gas ratios, and dust characteristics at high redshift, we introduce an additional parameter, $f_{\rm dust}$, which changes the total mass of the dust but does not change how the dust-to-gas ratio scales with metallicity or its spatial distribution in the galaxy.  This gives an idea of how little dust we actually need to create a Balmer break consistent with that of JD1, where there are only upper limits on dust mass.

In Figure~\ref{JD1}, we show the SED for $f_{\rm dust}=0,\ 1/20,\ 1/10,\ 1/3,\ 1/2,\ \&\  1$.  For $f_{\rm dust}=0$, the model that assumes no dust, the SED of AD3 is completely inconsistent with JD1.  The SED is extremely blue and dominated by the luminosity from young stellar populations.  However, for all other models that assume $f_{\rm dust}>0$, the SED of AD3 can be made to be appear reasonably consistent with JD1 indicating that very little dust is actually needed for the simulated galaxy to have an SED that is consistent with JD1.  In general the absolute value of $\chi^2$ for the AD3 fits to the JD1 photometry are extremely poor due to the very small error bars on the photometry.  AD3 is clearly not the exact same galaxy as JD1 so this is not particularly worrying as we only aim to show the general shape of the AD3 SED is qualitatively consistent with that of JD1 and that dust plays an extremely important role in reconciling the intrinsic continuum SED of AD3 with the photometry of JD1.   

\begin{figure}
\centerline{\includegraphics[scale=1]{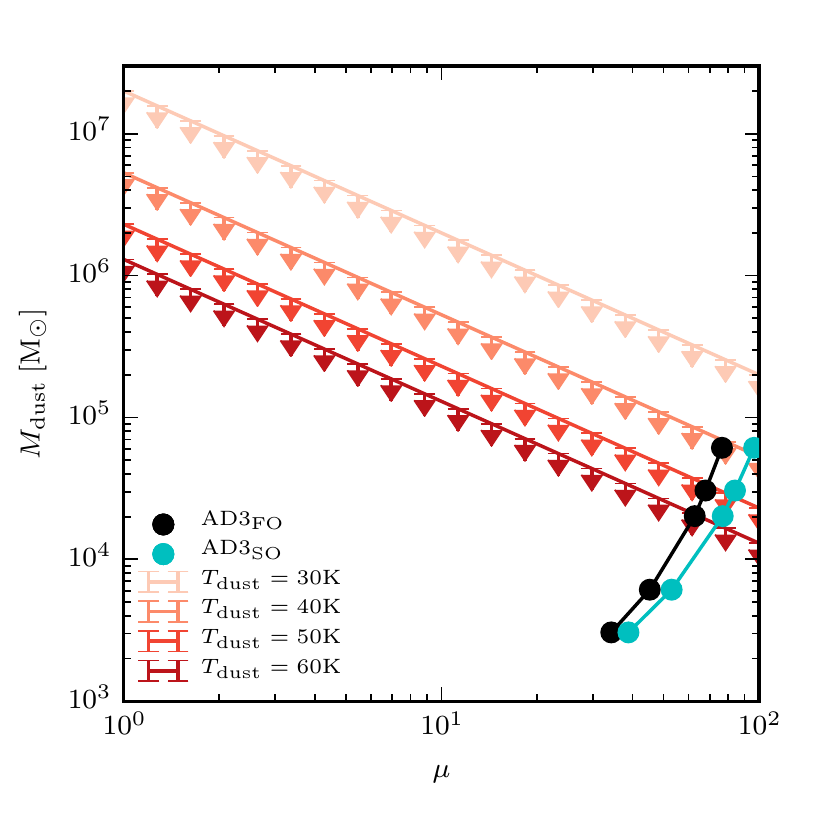}}
\caption{Dust mass as a function of magnification for JD1 for different assumptions on the dust temperature.  All lines show the $3\sigma$ upper limits.  The dust mass was estimated using a modified blackbody for the dust emission and an emissivity index of $\beta_{\rm d}=1.5$ using a model for the dust mass absorption coefficient as a function of frequency $\kappa=\kappa_0(\nu/\nu_0)^{\beta_{\rm d}}$, where $\kappa_0=10{\rm cm^2\ g^{-1}}$ at $250\mu$m. The black and cyan lines represent the dust masses of AD3 at $z=9.2$ as a function of magnification needed to fit the JD1 photometry for the face-on and side-on viewing angle, respectively.  The different points indicate the different dust-to-gas ratio normalisations shown in Figure~\ref{JD1}.  In all cases, the AD3 dust mass is in agreement with the observations for dust temperatures less than 40K.  For higher dust temperatures, a lower dust-to-gas ratio must be assumed.}
\label{JD1_dm}
\end{figure}

We can check whether the dust mass of AD3 for the different values of $f_{\rm dust}$ and magnification are consistent with the upper limits from JD1.  In Figure~\ref{JD1_dm}, we show the $3\sigma$ upper limits on the dust mass of JD1 as a function of the magnification of the object for different assumptions on the dust temperature.  If the dust temperature in JD1 is $<40$K, regardless of our assumed value for $f_{\rm dust}$, the dust mass in AD3 is consistent with the $3\sigma$ upper limits from the non-detection.  For higher dust temperatures, lower values of $f_{\rm dust}$ are required to make the simulations compatible with observations.  However, for $f_{\rm dust}=0.33$, which provides a qualitatively reasonable fit to the JD1 photometry, the dust mass of AD3 is consistent with all of the upper limits on the JD1 dust mass for dust temperatures $<60$K, which is the highest dust temperature assumed in \cite{Hashimoto2018}\footnote{Note that in the local Universe, dust temperatures $>60$K are only generally observed in ULIRGs hosting AGN \citep{Devriendt1999}.}.  This value is well within the uncertainties on the stellar yields, dust-to-gas ratios, dust properties, and dust production and destruction mechanisms.

In Table~\ref{JD1comp}, we compare the observed and inferred properties of JD1 with those of AD3 for the three dust models that provide the best fits to the JD1 photometry (i.e. $f_{\rm dust}=0.33,\ 0.50,\ \&\ 1.0$).  In Table~\ref{JD1comp}, we have adjusted the properties of JD1 for the assumed magnification factor of AD3 that gives the best fit for each dust model.  Note that in all cases, the magnification we must assume for AD3 is consistent within the uncertainties of the magnification of JD1.  In general, we find that the estimated SFR of JD1 is $3-4$ times lower for JD1 than what we measure in AD3.  This is not particularly surprising because in order to fit the photometry of JD1, \cite{Hashimoto2018} assumed that there was no dust attenuation for the young stellar population in their two-component model.  In contrast, AD3 has considerably more attenuation which preferentially obscures the young stars and hence the SFR of AD3 must be higher than that assumed for JD1 when the SED is fit without dust.  Similarly, the stellar mass estimated for JD1 is also lower by a factor of $2-3$ for the same reason.  The metallicity that we measure in AD3 is entirely consistent with that estimated for JD1, although there are indeed large uncertainties.  As stated before, if we assume that the dust temperature in JD1 is close to the CMB temperature at $z\sim9.1$, the dust masses we find for AD3 are a factor of $5-10$ below the $3\sigma$ upper limits for JD1.

\begin{table*}
\centering
\begin{tabular}{@{}lcccccc@{}}
\hline
Property & JD1 & AD3 & JD1 & AD3 & JD1 & AD3 \\
 & $\mu=69.8$ & $f_{\rm dust}=0.33$ & $\mu=75.9$ & $f_{\rm dust}=0.50$ & $\mu=86.5$ & $f_{\rm dust}=1.0$ \\ 

\hline
\hline
SFR (${\rm M_{\odot}yr^{-1}}$) & $0.60^{+0.11}_{-0.16}$ & 2.05 & $0.55^{+0.11}_{-0.14}$ & 2.05 & $0.49^{+0.09}_{-0.11}$ & 2.05\\
Stellar Mass ($10^8{\rm M_{\odot}}$) & $1.55^{+0.76}_{-0.26}$ & 3.81 & $1.42^{+0.70}_{-0.24}$ & 3.81 & $1.25^{+0.61}_{-0.21}$ & 3.81 \\
Metallicity & $0.004\pm0.004$ & 0.003 & $0.004\pm0.004$ & 0.003 & $0.004\pm0.004$ & 0.003\\
30K Dust Mass ($10^5{\rm M_{\odot}}$) & $<2.86$ & 0.611 &  $<2.63$ & 0.306 &  $<2.31$ & 0.204\\ 
$[{\rm OIII}]_{88\mu}$ Luminosity ($10^7L_{\odot}$) & $1.06\pm0.23$ & 0.096$Z_O/Z_{\rm gal}$ & $0.97\pm0.21$ & 0.096$Z_O/Z_{\rm gal}$ & $0.86\pm0.18$ & 0.096$Z_O/Z_{\rm gal}$\\
$[{\rm CII}]_{158\mu}$ Luminosity ($10^7L_{\odot}$) & - & 1.870$Z_C/Z_{\rm gal}$ & - & 1.870$Z_C/Z_{\rm gal}$ & - & 1.870$Z_C/Z_{\rm gal}$\\
$[{\rm OIII}]_{5007{\rm \AA}}$ Luminosity ($10^7L_{\odot}$) & - & 0.793$Z_O/Z_{\rm gal}$ & - & 0.793$Z_O/Z_{\rm gal}$ & - & 0.793$Z_O/Z_{\rm gal}$ \\
Ly$\alpha$ Luminosity ($10^7L_{\odot}$) & $1.78\pm0.46$ & - & $1.63\pm0.42$ & - & $1.43\pm0.37$ & - \\
\hline
\end{tabular}
\caption{Galaxy properties as a function of redshift.  All masses are quoted in $\log_{10}({\rm M/M_{\odot}})$.  Metallicities represent the mean mass-weighted gas-phase metallicity, while the SFRs are in units of ${\rm M_{\odot}/yr}$ averaged over the previous 10Myr.}
\label{JD1comp}
\end{table*}

\subsubsection{Emission Line Properties}
\label{lineemission}
JD1 was also observed by ALMA where the far-infrared [OIII]~$88\mu$m line luminosity has been constrained.  Upcoming observations with ALMA and potentially JWST will likely also constrain the luminosities of the [CII]~$158\mu$m fine-structure line and the [OIII]~$5007\AA$ nebular emission line.  Since our simulations contain the density, temperature, metallicity, and inhomogeneous radiation field for each cell, we can post-process the simulations with a photoionisation code (i.e. CLOUDY) to obtain the emission line luminosities for AD3 (see also \citealt{Moriwaki2018} for high-redshift [OIII] luminosity predictions from simulations).  Details of our method are presented in Appendix~\ref{methods}.

Our method predicts that the [OIII]~$88\mu$m luminosity of AD3 is about a factor of 10 smaller than that for JD1.  We expect some differences in the luminosity due to the fact that they are indeed different objects; however, a factor of 10 may seem excessive if they are indeed very similar.  \cite{Steidel2016} have shown that highly super solar [O/Fe] ratios are expected for $z\simeq2-3$ star-forming galaxies which would naturally occur if the dominant metal enrichment channel is via core-collapse SN.  For these same galaxies, they find that while the stellar metallicity is $\sim0.1Z_{\odot}$, which is very similar to that of AD3, the nebular oxygen abundance must be $\sim0.5Z_{\odot}$.  This gas phase is not always explicitly resolved by our simulation.  Hence, if we apply their factor 5 correction to our luminosities, we find that our [OIII]~$88\mu$m luminosities are within a factor of $\sim2$ of JD1 which is in much better agreement.  Because our systems are at such a high redshift, we expect that core-collapse SN are the dominant enrichment mechanism as it is unlikely that type-Ia SN would have had time to occur. We have added a factor of $Z/Z_{\rm gal}$ to all our metal line luminosity predictions in Table~\ref{JD1comp} to emphasise that they should be rescaled based on the expected elemental abundance in the [OIII] emitting regions with respect to the global metallicity of our galaxy that assumes solar abundance ratios.

Although [CII]~$158\mu$m and [OIII]~$5007\AA$ have not yet been observed for JD1, we can make predictions for their expected luminosities based on AD3.  Assuming the same factor 5 boost in oxygen abundance, we would expect that the [OIII]~$5007\AA$ luminosity is within a factor of $\sim2$ of $\sim4\times10^7L_{\odot}Z_O/5Z_{\rm gal}$ while the [CII]~$158\mu$m luminosity is within a factor of a few times 1.870$\times10^7L_{\odot}Z_C/Z_{\rm gal}$.  Such magnitude [CII]~$158\mu$m luminosities have already been detected in $z>7.5$ objects \citep[e.g.][]{Knudsen2017}.

The final line luminosity that has been constrained for JD1 is Ly$\alpha$.  This line is much more difficult to model because it is resonant and thus requires Monte-Carlo radiation transfer to simulate how the photons diffuse both spatially and in frequency.  Interestingly, the Ly$\alpha$ line is blue-shifted with respect to the [OIII]~$88\mu$m.  This could, for instance, indicate that we are seeing an inflow; but this would require a massive ionised bubble around the source such that the Ly$\alpha$ isn't absorbed by the IGM.  Alternatively, it could be the case that neither line represents the true systemic redshift of the object.  In our simulations, [OIII]~$88\mu$m probes the hot gas around star forming regions and thus the line may be picking up the local velocities of the individual star forming regions.  Because of the compact nature of high-redshift objects, these velocities can be hundreds of km/s offset from systemic.  Due to the intricacies in modelling Ly$\alpha$, we leave a more complete discussion of this line to future work.

\begin{figure*}
\centerline{\includegraphics[scale=1]{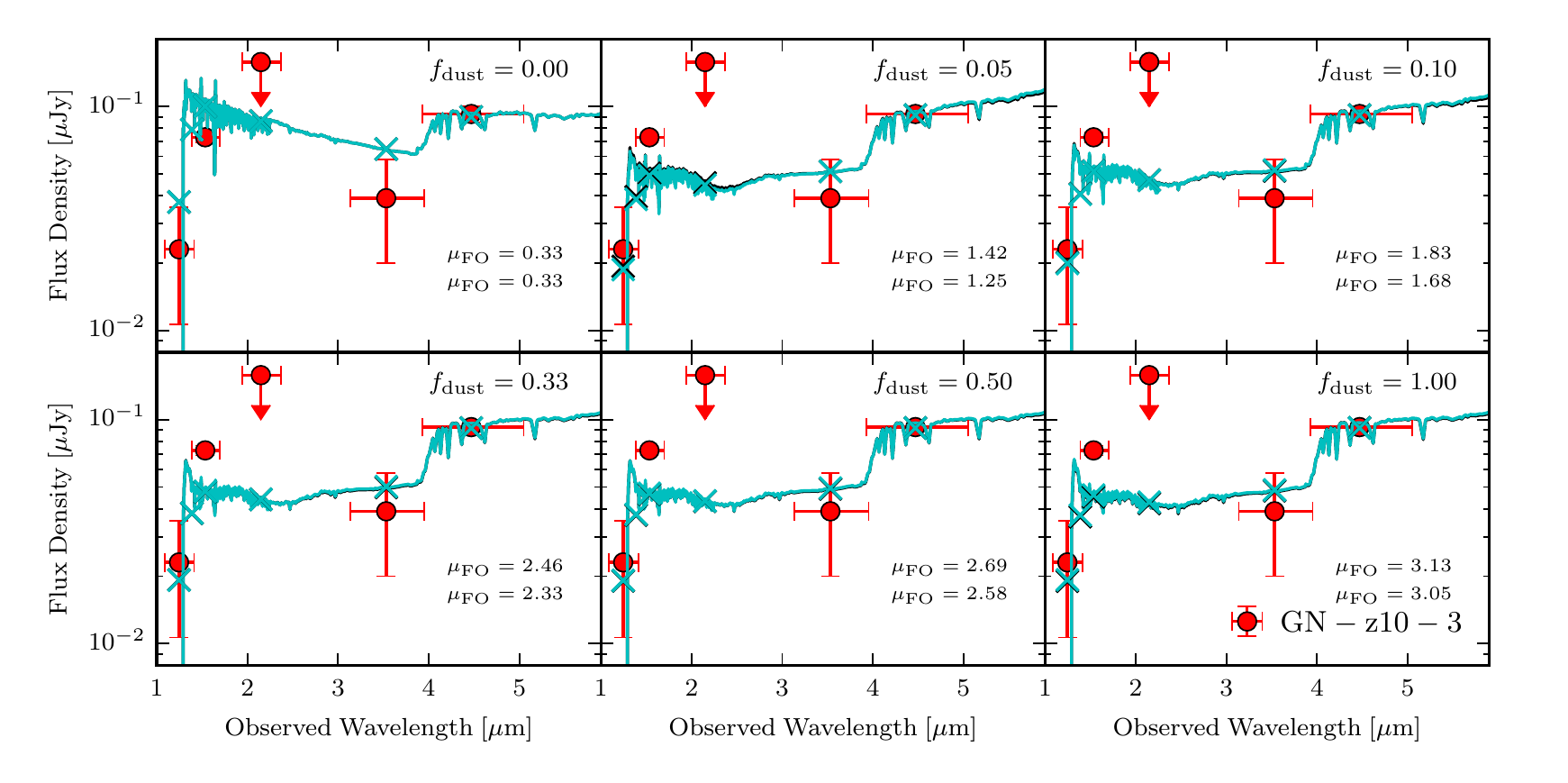}}
\caption{Photometry of GN-z10-3 compared to the $z=9.4$ snapshot of AD1 for two viewing angles.  All lines and points are the same as in Figure~\ref{JD1}.}
\label{GN-10-3}
\end{figure*}
\begin{figure*}
\centerline{\includegraphics[scale=1]{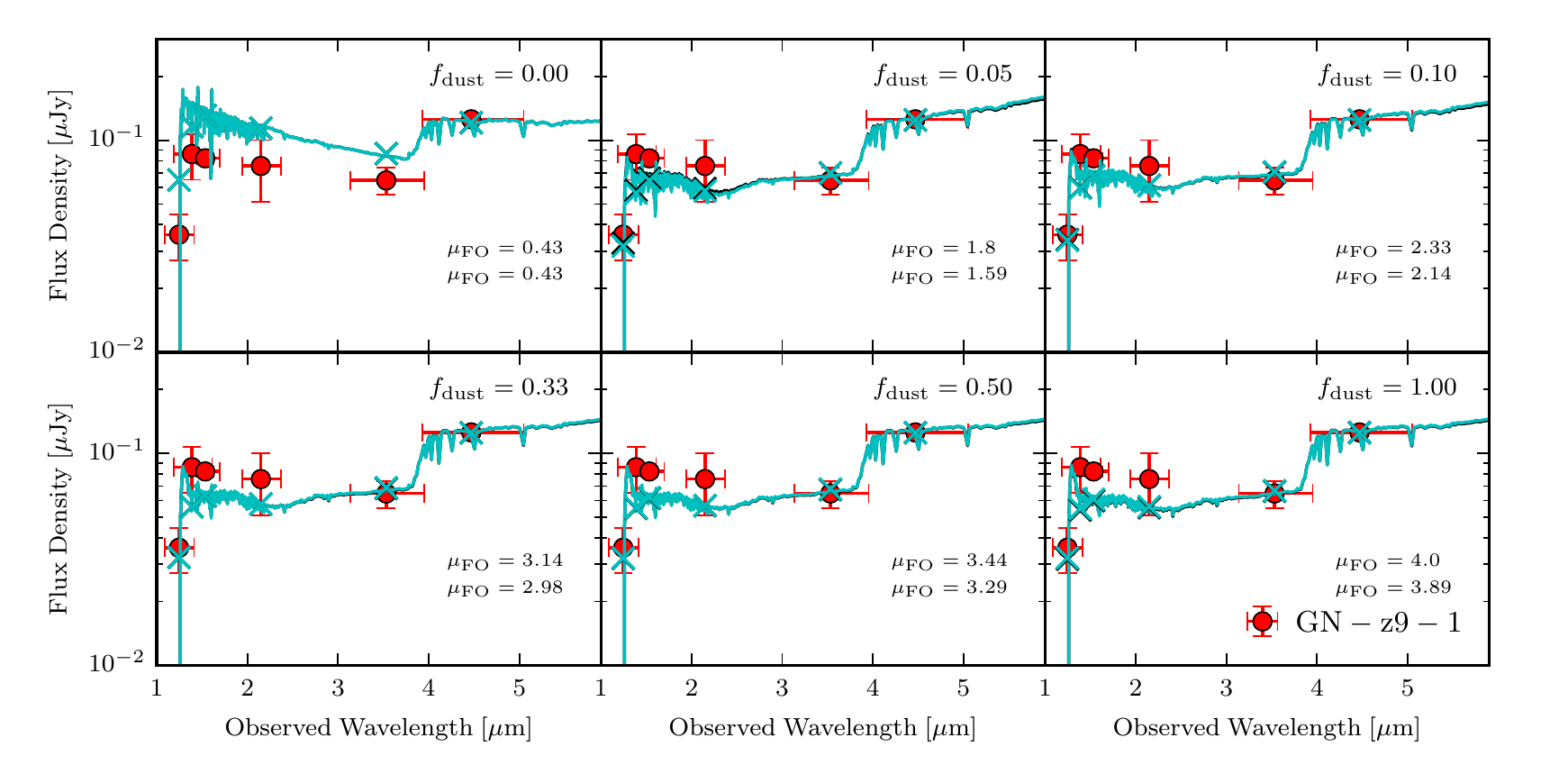}}
\caption{Photometry of GN-z9-1 compared to the $z=9.4$ snapshot of AD1 for two viewing angles.  All lines and points are the same as in Figure~\ref{JD1}.}
\label{GN-z9-1}
\end{figure*}

In summary, a Balmer break with the same strength as that in JD1 can be easily reproduced by one of our three simulated galaxies at the same redshift, indicating that objects like JD1 might not be uncommon at these redshifts.  Even though the dust content in simulations is highly uncertain, we show that our simulated SEDs are reasonably consistent with the photometry from JD1, even when it is modulated down by a factor of $\sim10$ because the dust is preferentially located near young star-forming regions.  If AD3 is indeed a JD1 analog, it requires a large magnification factor, a significantly super-solar solar oxygen abundance, and the [CII]~$158\mu$m and [OIII]~$5007\AA$ should be observable by ALMA and JWST, respectively.  Furthermore, the star formation rate history of this object would not be exponentially declining, but rather have a peak at $\sim350$Myr after the Big Bang, and remain bursty, but on average decline by a factor of a few ever since.

\subsection{GN-z10-3 \& GN-z9-1}
Although our simulated galaxies can broadly match the properties of JD1 which provides insight into the ISM, dust properties, and star formation history of the galaxy, there remains a large uncertainty on the magnification of the object.  It remains an open question of whether or not JD1 is a unique object, or if strong Balmer breaks are commonplace at high redshifts.  Despite having only photometric redshifts, if GN-z10-3 and GN-z9-1 are confirmed to be at their maximum likelihood photometric redshift, they will also be in a regime where their IRAC channel 2 excess is caused by a Balmer break rather than nebular emission lines.  For this reason, we compare whether any of our simulated SEDs can reproduce the potential Balmer breaks observed for these galaxies.  

In Figures~\ref{GN-10-3}~\&~\ref{GN-z9-1}, we compare the photometry of GN-z10-3 and GN-z9-1 to the $z=9.4$ snapshot of AD1. We have redshifted the AD1 spectra to the maximum likelihood redshift for each object and show the SED for the six different values of $f_{\rm dust}$.  Furthermore, we have renormalised the SED of AD1 assuming a constant scale factor so that the flux of our simulated galaxy in IRAC channel 2 matches the observations, so that the strengths of the Balmer breaks can be compared.  If the IRAC channel 2 excess in GN-z10-3 is due to a Balmer break, our simulated galaxies would have a difficult time reproducing its strength.  The most consistent AD1 models can fit the Hubble and IRAC channel 2 fluxes but reside towards the upper limit on the $1\sigma$ error bar in the IRAC channel 1 band.  The error bars on the flux in this band are large so it could be argued that our simulated galaxy is formally consistent with this measurement; however, if the centroid of this distribution is confirmed, our simulated SEDs would be inconsistent with this galaxy.  In order to fit the photometry of GN-z10-3, we must also assume a stellar mass that is $\sim2-3$ times larger (depending on dust mass) than that of AD1 at $z=9.4$, which gives a stellar mass estimate of $\sim7-10\times10^9{\rm M_{\odot}}$.  This stellar mass could be lowered if the dust in AD1 was distributed more like it is in AD3, where there is a much stronger preference for the dust to be around young stars (see Figure~\ref{tau_dust}).

In contrast to GN-z10-3, the Balmer break, although present, is weaker for GN-z9-1.  In Figure~\ref{GN-z9-1}, we show that the $z=9.4$ SED of AD1 appears to be in much better agreement with this galaxy than for GN-z10-3.  The agreement once again relies on dust.  In the top left panel of Figure~\ref{GN-z9-1}, one can see that without dust, the SED of AD1 is much too blue to match the photometry of GN-z9-1.  Even with a very small amount of dust (i.e. $f_{\rm dust}=0.05$), the shape of the AD1 SED is in very good agreement with the photometry of GN-z9-1.  In order to match the normalisation, depending on dust content, the stellar mass of GN-z9-1 must be $\sim2-4$ times that of AD1.  Once again if the dust distribution in GN-z9-1 was slightly different to AD1, this number may decrease.  Regardless of the exact normalisation, it is clear that this potential Balmer break is naturally reproduced by our simulations.  We argue that these types of objects are not weird or abnormal, rather they can form in a hierarchical structure formation scenario, within a $\Lambda$CDM cosmology, as long as dust is accounted for.

\section{Discussion and Conclusions}
We have compared the SEDs of galaxies in a high-resolution cosmological radiation hydrodynamics simulations with the photometry of three $z\sim9$ galaxies (one spectroscopically confirmed and two with photometric redshifts), MACS~1149\_JD1, GN-z10-3, and GN-z9-1, to determine whether the observed Balmer breaks can be reproduced by simulations.  The observed systems were specifically selected so that their IRAC~channel~2 excess is not contaminated by the [OIII] and H$\beta$ nebular emission lines.  Understanding this excess is key for determining the assembly history of high-redshift galaxies as well as the onset of cosmic dawn.  However, even for our very high redshift simulated systems, dust is a crucial component of the SED modelling.  Our main conclusions are as follows:

\begin{itemize}
\item  For two of our three most massive simulated systems, the dust resides preferentially around the young star-forming regions, thus enhancing the presence of the Balmer break, and complicating its interpretation. 
\item The strength of the Balmer break in JD1 can be easily reproduced by one of our three simulated galaxies, AD3, indicating that objects like JD1 may be relatively normal at high redshifts.  
\item If AD3 is a direct analog of JD1, we suggest that it has a stellar mass of $~\sim1.25\times10^8$M$_{\odot}$, a SFR of $0.49$M$_{\odot}$yr$^{-1}$, a relatively large magnification (between 20-90), a dust mass $<10^5$M$_{\odot}$, and a super-solar ratio of oxygen abundance with respect to iron.  Star formation in the object likely occurred within 200Myr of the Big Bang.
\item The photometry of GN-z10-3 and GN-z9-1 can be reasonably reproduced by the SED of AD1 but we require that the stellar mass of the objects be a factor of $\sim2-7$ times larger than that of AD1 (which has a stellar mass of $10^{9.54}$M$_{\odot}$ at $z=9.4$).
\item Objects with very similar instantaneous stellar masses and SFRs (i.e. AD2 and AD3) can have extremely different dust morphologies, leading to strong variations in SED shape and attenuation.  Interestingly, the objects that begin forming stars first (i.e. AD2) may not exhibit Balmer breaks as strong as those that form stars later (i.e. AD3).
\end{itemize}

Future observations that directly observe the dust emission from objects like JD1 may help to constrain the dust content of these galaxies to determine whether the systems that exhibit strong Balmer breaks are consistent with the simulated objects.

Directly comparing our simulations with high-redshift observations is extremely non-trivial and certain caveats should be considered when interpreting our work.  Firstly, there remain little constraints on the dust production and destruction mechanisms, evolution, and properties at these extreme redshifts.  Furthermore, we have assumed that the dust mass scales with metallicity following the relations from \citep{Remy2014} that have considerable scatter.   If the dust distribution is not completely coupled to the metal distribution, the dust optical depths seen by individual stars will be different from what is presented.  

We have only considered the continuum SEDs of these objects without any nebular emission lines.  However, for old stellar populations, which are likely to dominate the SEDs of galaxies with large Balmer breaks, we expect that the nebular lines are extremely sub-dominant compared to the stellar and nebular continuum.  Likewise, we have only considered attenuation from inside the galaxy rather than scattering, or attenuation and scattering from material in the IGM.  This would require running radiation transfer in a full light cone which is beyond the scope of this work. 

With these caveats in mind, even with little amounts of dust (i.e. $<10^5$M$_{\odot}$) present in these high-redshift objects, as long as it is preferentially located around young star-forming regions, our primary conclusions are likely robust.

\section*{Acknowledgements}
We acknowledge useful discussions with Pascal Oesch who drew our attention to the objects GN-z10-3 and GN-z9-1.  We also thank Jeremy Blaizot, Akio Inoue, Tayson Kimm, Clotilde Laigle, Joki Rosdahl, and Andrea Ferrara for discussions regarding content in this manuscript.  NL and RSE acknowledge funding from the European Research Council (ERC) under the European Union's Horizon 2020 research and innovation programme (grant agreement No. 669253).   H.K. thanks the Beecroft fellowship, the Nicholas Kurti Junior Fellowship, and Brasenose College. The research of AS and JD is partially funded by the generosity of Adrian Beecroft.

This work was performed using the DiRAC Data Intensive service at Leicester, operated by the University of Leicester IT Services, which forms part of the STFC DiRAC HPC Facility (www.dirac.ac.uk). The equipment was funded by BEIS capital funding via STFC capital grants ST/K000373/1 and ST/R002363/1 and STFC DiRAC Operations grant ST/R001014/1. DiRAC is part of the National e-Infrastructure.




\bibliographystyle{mnras}
\bibliography{biblio} 




\appendix
\section{Nebular Continuum Emission}
\label{methods}
In this Appendix we describe our method for measuring nebular continuum emission.  We can exploit the fact that we resolve both a multiphase ISM and an inhomogeneous radiation field within each cell of our simulation to measure the nebular continuum emission on a cell-by-cell basis within the simulation.  To do this, we employ the spectral synthesis code {\small CLOUDY} \citep{Ferland2017}.  We have run {\small CLOUDY} on 850,000 simulation cells from the central region of AD1 at $z=10$.  A slab of gas is set up in an open geometry at a large distance from the central source.  The depth of the slab is set to the length of the cell in the simulation and we assume a constant temperature, metallicity, and density to be consistent with the simulation.  The slab is irradiated with a flux and spectral shape consistent with the simulation.  We also assume an isotropic background from the CMB.  For each model, we extract the nebular continuum using the results from the {\small save continuum} command between rest-frame wavelengths of 900$\AA$-8500$\AA$ which more than covers the range in wavelength that we are interested in.  

\begin{figure}
\centerline{\includegraphics[scale=1]{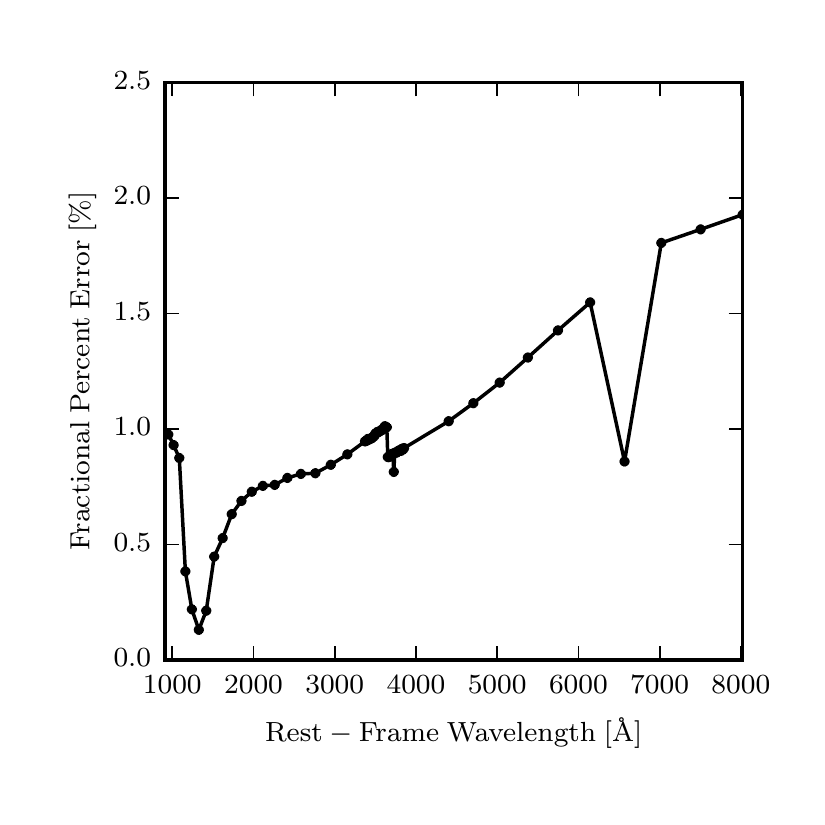}}
\caption{Fractional percentage error ($|L_{\rm Prediction}-L_{\rm CLOUDY}|/L_{\rm CLOUDY}$) in the total luminosity of nebular continuum emission on the test data set.  The points represent 70 locations where we predict the nebular continuum.  The nebular continuum is sampled much better at the locations near the Balmer Jump to better capture this feature.  In all cases, we predict the total nebular continuum luminosity to better than 2\% accuracy.}
\label{neberr}
\end{figure}

It is currently computationally impractical to run {\small CLOUDY} models for the hundreds of millions of cells in our simulation in each output and our parameter space is too large to effectively create a grid of models.  Thus, in order to calculate the nebular continuum for all other cells in the simulation, we have trained a Random Forest \citep{Ho1995,Breiman2001,Geurts2006}, an ensemble machine learning method, to use the density, temperature, metallicity, and radiation field within a simulation cell to predict what the value of the nebular continuum will be at 70 different wavelengths in the range described earlier.  The sampling of the points across the continuum is uneven as we increase the resolution around the Balmer Jump to make sure that this feature is not smoothed out.  To assess the accuracy of our method, the Random Forest is only trained on a subset of the data (85\%) while the remaining data is reserved to assess the accuracy of the method.  Within the training set, we perform a cross validation to determine the optimum number of trees in the forest (in our case we use 100).  In Figure~\ref{neberr}, we show the percentage error difference in the prediction of the total nebular continuum luminosity at the 70 different wavelengths for the data that was not used to train the algorithm.  At all wavelengths, the total luminosity that we estimate using the Random Forest is within 2\% of that generated with {\small CLOUDY} indicating that we can obtain a very accurate continuum luminosity for a simulated galaxy using our method.  Furthermore, our method is orders of magnitude faster for generating nebular continuum emission for every cell compared to using {\small CLOUDY}.  To generate the total nebular continuum emission for each galaxy, we sum the results for all cells within the virial radius and perform the same dust attenuation calculation that is described earlier.

A similar method is used to generate the IR land nebular in luminosities discussed in Section~\ref{lineemission}; however, in this case, the Random Forest is trained on the line emission rather than the continuum emission.

Note that our method for calculating nebular continuum emission is significantly different from others in the literature.  For example, \cite{Wilkins2016} post-processed the BlueTides simulation assuming a fixed gas density of 100~${\rm cm^{-3}}$ and only perform the calculation around star particles where the metallicity of the gas is set to that of the star.  We prefer to exploit the inhomogeneous gas and radiation field conditions throughout our simulation to estimate this emission as this may impact the results.  \cite{Wilkins2016} found that nebular emission can play a significant role in modulating the shape and normalization of the SED.  We can also confirm that under certain circumstances, the nebular continuum emission can contribute upwards of 50\% of the total continuum luminosity of a galaxy.  However, we note that for the $z=9.2$ snapshot of AD3 which is most similar to JD1, the nebular continuum makes a subdominant contribution to the total SED compared to the stars.

\bsp	
\label{lastpage}
\end{document}